\pdfoutput=1
\documentclass{acm_proc_article-sp}

\usepackage{url}

\begin{document}

\title{PRAGMA-ENT: Exposing SDN Concepts to Domain Scientists in the Pacific Rim}

\numberofauthors{16} %  in this sample file, there are a *total*
\author{
\alignauthor
Kohei Ichikawa\\
       \affaddr{Nara Institute of Science and Technology}\\
       \affaddr{8916-5 takayama-cho, Ikoma, Japan}\\
       \email{ichikawa@is.naist.jp}
\alignauthor
Mauricio Tsugawa\\
       \affaddr{University of Florida}\\
       \affaddr{968 Center Dr, Gainesville, FL, US}\\
       \email{tsugawa@acis.ufl.edu}
\alignauthor
Jason Haga\\
       \affaddr{National Institute of Advanced Industrial Science and Technology}\\
       \affaddr{1-1-1 Umezono, Tsukuba, Japan}\\
       \email{jh.haga@aist.go.jp}
\and
\alignauthor
Hiroaki Yamanaka\\
       \affaddr{National Institute of Information and Communications Technology}\\
       \affaddr{4-2-1, Nukui-Kitamachi, Koganei, Japan}\\
       \email{hyamanaka@nict.go.jp}
\alignauthor
Te-Lung Liu\\
       \affaddr{National Applied Research Laboratories}\\
       \affaddr{No. 7, R\&D 6th Rd., Hsinchu Science Park, Hsinchu, Taiwan}\\
       \email{tlliu@narlabs.org.tw}
\alignauthor
Yoshiyuki Kido\\
        \affaddr{Osaka University}\\
        \affaddr{1-1 Yamadaoka, Suita, Japan}\\
        \email{kido@cmc.osaka-u.ac.jp}
%% 
%% Following authors are added as \additionalauthors
%% \alignauthor
%% Pongsakorn U-chupala\\
%%        \affaddr{Nara Institute of Science and Technology, Japan}\\
%%        \email{pongsakorn.uchupala.pm7\\@is.naist.jp}
%% \alignauthor
%% Che Huang\\
%%        \affaddr{Nara Institute of Science and Technology, Japan}\\
%%        \email{huangche@is.naist.jp}
%% \alignauthor
%% Chawanat Nakasan\\
%%        \affaddr{Nara Institute of Science and Technology, Japan}\\
%%        \email{chawanat.nakasan.cb5\\@is.naist.jp}
%% \and 
%% \alignauthor
%% Susumu Date\\
%%        \affaddr{Osaka University, Japan}\\
%%        \email{date@cmc.osaka-u.ac.jp}
%% \and
%% \alignauthor
%% Philip Papadopoulos\\
%%        \affaddr{University of California,\\ San Diego, US}\\
%%        \email{ppapadopoulos\\@mail.ucsd.edu}
%% \alignauthor
%% Jose Fortes\\
%%        \affaddr{University of Florida, US}\\
%%        \email{fortes@acis.ufl.edu}
}
\additionalauthors{Additional authors:
Pongsakorn U-chupala (Nara Institute of Science and Technology, Japan, email: pongsakorn.{\allowbreak}uchupala.{\allowbreak}pm7{\allowbreak}@is.naist.jp), 
Che Huang (Nara Institute of Science and Technology, Japan, email: huangche@is.naist.jp), 
Chawanat Nakasan (Nara Institute of Science and Technology, Japan, email: chawanat.nakasan.cb5@is.naist.jp), 
Jo-Yu Chang (National Applied Research Laboratories, Taiwan, email: stoca{\allowbreak}@narlabs.org.tw),
Li-Chi Ku (National Applied Research Laboratories, Taiwan, email: lku@narlabs.org.tw),
Whey-Fone Tsai (National Applied Research Laboratories, Taiwan, email: wftsai@nchc.org.tw),
Susumu Date (Osaka University, Japan, email: date@cmc.osaka-u.ac.jp),
Shinji Shimojo (Osaka University, Japan, email: shimojo@cmc.osaka-u.ac.jp),
Philip Papadopoulos (University of California, San Diego, US, email: ppapadopoulos@mail.ucsd.edu)
and
Jose Fortes (University of Florida, US, email: fortes@acis.ufl.edu).
}

\date{15 August 2015}

\maketitle
\begin{abstract}
The Pacific Rim Application and Grid Middleware Assembly (PRAGMA) is an international community of researchers that actively collaborate to address problems and challenges of common interest in eScience. The PRAGMA Experimental Network Testbed (PRAGMA-ENT) was established with the goal of constructing an international software-defined network (SDN) testbed to offer the necessary networking support to the PRAGMA cyberinfrastructure. PRAGMA-ENT is isolated, and PRAGMA researchers have complete freedom to access network resources to develop, experiment, and evaluate new ideas without the concerns of interfering with production networks.
 
In the first phase, PRAGMA-ENT focused on establishing an international L2 backbone. With support from the Florida Lambda Rail (FLR), Internet2, PacificWave, JGN-X, and TWAREN, PRAGMA-ENT backbone connects Open\-Flow-enabled switches at University of Florida (UF), University of California San Diego (UCSD), Nara Institute of Science and Technology (NAIST, Japan), Osaka University (Japan), National Institute of Advanced Industrial Science and Technology (AIST, Japan), and National Center for High-Performance Computing (Taiwan).
 
The second phase of PRAGMA-ENT consisted of evaluation of technologies for the control plane that enables multiple experiments (i.e., OpenFlow controllers) to co-exist. Preliminary experiments with FlowVisor revealed some limitations leading to the development of a new approach, called AutoVFlow.
This paper will share our experience in the establishment of PRAGMA-ENT backbone (with international L2 links), its current status, and control plane plans. Discussion on preliminary application ideas, including optimization of routing control; multipath routing control; and remote visualization will also be discussed.
\end{abstract}

\category{C.2.1}{Network Architecture and Design}{Distributed networks}
\category{C.2.4}{Distributed Systems}{Distributed applications}

\terms{Experimentation, Design, Management, Performance, Mesurement}

% \keywords{ACM proceedings, \LaTeX, text tagging} % NOT required for Proceedings

\section{Introduction}
Effective sharable cyberinfrastructure is fundamental for collaborative research in eScience. There are many projects to build large-scale cyberinfrastructure, including TeraGrid \cite{catlett2006teragrid}, Open Science Grid \cite{pordes2007open}, FutureSystems \cite{von2010design} and so on. The PRAGMA testbed is also one of the cyberinfrastructure built by the Pacific Rim Application and Grid Middleware Assembly (PRAGMA - \url{http://www.pragma-grid.net}) \cite{Arzberger2008,tanaka2013building}. PRAGMA is an international community of researchers that actively collaborate to address problems and challenges of common interest in eScience. The goal of the PRAGMA testbed is to enable researchers to effectively use cyberinfrastructure for their collaborative work. For this purpose, the participating researchers mutually agree to deploy a reasonable and minimal set of computing resources to build the testbed. The original fundamental concept of the PRAGMA testbed was based on Grid computing technologies, but recently, the focus of the community has shifted to a virtualized infrastructure (or Cloud computing) because virtualization technology provides more dynamic and flexible resources for collaborative research.

With this move to virtualized infrastructures, the networking testbed is also becoming a major concern in the community. In the virtualized infrastructures, newly emerging virtual network technologies are essential to connect these resources to each other and the evaluation of these technologies requires a networking testbed to be built. National network projects in each country, such as JGN-X and Internet2, are attempting to provide Software Defined Network (SDN) services to meet such demands. However, adapting one of these national network services is not sufficient for evaluating global scale cyberinfrastructure like PRAGMA.

The PRAGMA-ENT expedition was established in October 2013 with the goal of constructing a breakable international SDN/OpenFlow testbed for use by PRAGMA researchers and collaborators. PRAGMA-ENT is breakable in the sense that it will offer complete freedom for researchers to access network resources to develop, experiment, and evaluate new ideas without concerns of interfering with a production network. PRAGMA-ENT will provide the necessary networking support to the PRAGMA multi-cloud and user-defined trust envelopes. This will expose SDN to the broader PRAGMA community and facilitate the long-tail of eScience by creating new collaborations and new infrastructure among institutes in the Pacific Rim area. In this paper, we will describe and discuss the PRAGMA-ENT architecture, deployment, and applications, with an emphasis on its relationship with the Pacific Rim region.

The rest of this paper is structured as follows. Section 2 describes previous research on other testbeds using SDN technologies. Section 3 explains our architecture and deployment of data plane and control plane for PRAGMA-ENT. Section 4 introduces our applications and experiments using PRAGMA-ENT. Section 5 concludes the paper and describes the future plan.

\section{Related work}
SDN concepts are a newly established paradigm that features the separation of the control plane from the data plane \cite{casado2014abstractions}. The control plane has various abstractions that enable administrators/users to control how information flows through the data plane \cite{rothenberg2014open} by exposing a programmable interface. One advantage of this is that the controllers have a global view of the infrastructure topology and network state, thus providing unprecedented control over the network, which increases the efficiency, extensibility, and security of the network. OpenFlow is an open standard protocol that enables the control plane to interact with the data plane \cite{mckeown2008openflow}.

Advanced Layer 2 Service (AL2S) offered by Internet2 is the most successful network service using SDN technologies \cite{al2s,jeff2013SDN}. AL2S provides a dynamic Layer 2 provisioning service across the Internet2 backbone network. Users of Internet2 can provision point-to-point VLANs to other Internet2 users using a Web portal for AL2S. The software talks to an OpenFlow controller to push OpenFlow rules into the networking devices and allocates a VLAN between users across the network backbone in a very dynamic manner. Since the OpenFlow rules perform any necessary VLAN translation on any of the networking nodes in the path, it eliminates the long provisioning delay for matching VLAN IDs over the WAN. This service is very useful to provision a dedicated network for collaborative research, but the AL2S itself does not provide an opportunity for the users to deploy their own OpenFlow controller. Thus, the users do not access and control each networking switch directly.

Research Infrastructure for large-Scale network Experiments (RISE) offered by JGN-X is a very unique network service that allows its users to deploy their own OpenFlow controller on the network testbed \cite{ishii2013study}; therefore, users can fully control the assigned testbed using their controller. RISE creates multiple virtual switch instances (VSI) on each switch device and assigns them for users. Each VSI acts as a logical OpenFlow switch and is assigned to the user controller. This design is ideal for the PRAGMA testbed since we can have our own OpenFlow controller for the testbed. Internet2 is also beginning a similar service, called network virtualization service (NVS) that allows users to bring their own controller for the testbed, but it is not open for users at this time. We therefore decided to use RISE services as our backbone, and develop our data plane by connecting our OpenFlow switches deployed at participating sites to the backbone.

\section{Architecture and Deployment}

\subsection{Data Plane}

\begin{figure}
	\centering
	\includegraphics[width=\linewidth]{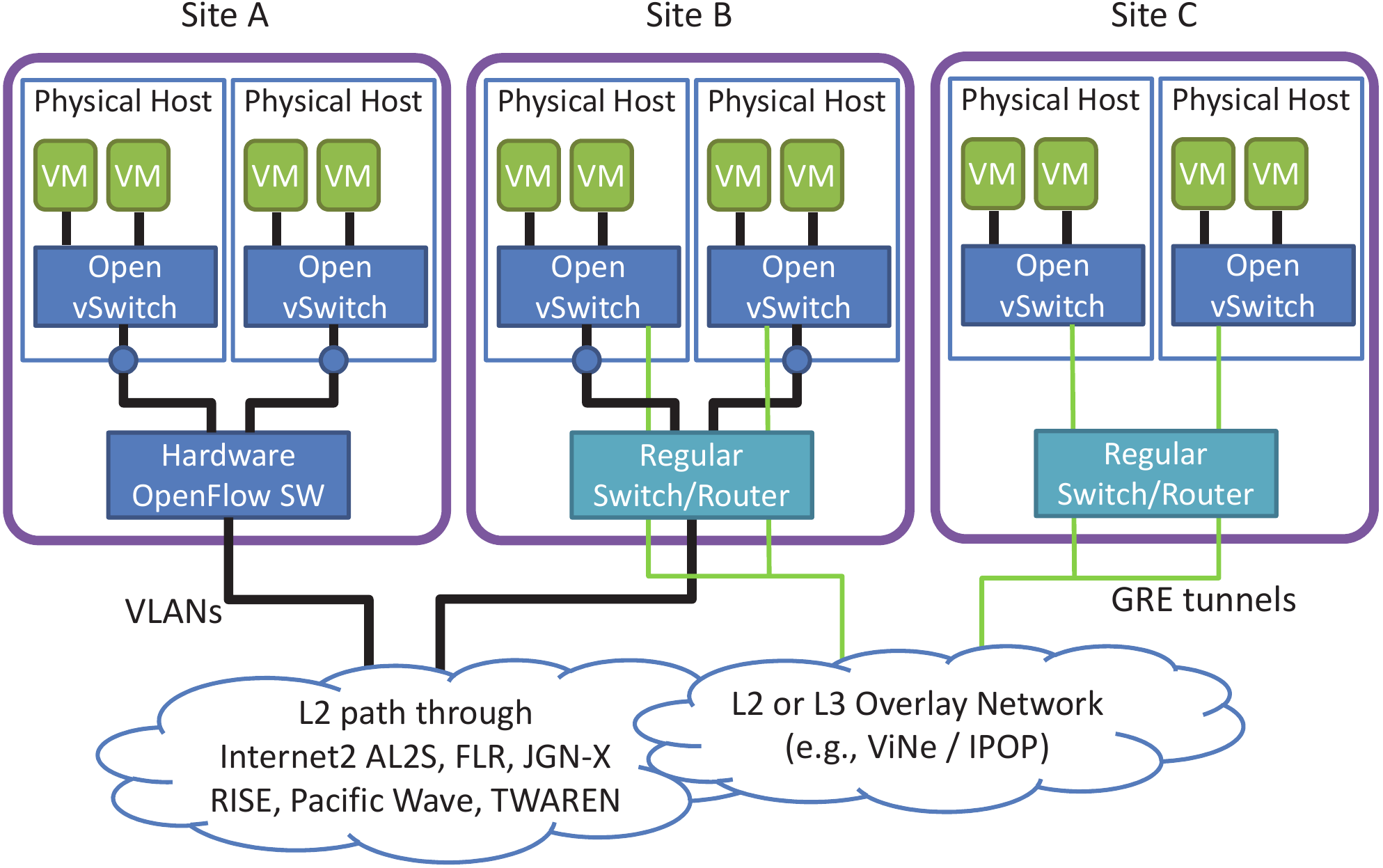}
	\caption{PRAGMA-ENT Data Plane. Hardware and software OpenFlow switches are connected via direct Wide-Area Layer-2 paths, Generic Routing Encapsulation (GRE) tunnels, and/or ViNe/IPOP overlays.}
	\label{fig:ent-data-plane}
\end{figure}

PRAGMA-ENT connects heterogeneous resources in different countries as illustrated in Figure \ref{fig:ent-data-plane}. Participating sites may or may not have OpenFlow-enabled switches, and may or may not have direct connection to PRAGMA-ENT Layer-2 backbone. For participants not on the PRAGMA-ENT Layer-2 backbone, overlay network technologies (such as ViNe \cite{tsugawa2006virtual} and IPOP \cite{IPOP}, developed at UF) will offer the necessary connectivity extensions.

\begin{figure}
	\centering
	\includegraphics[width=\linewidth]{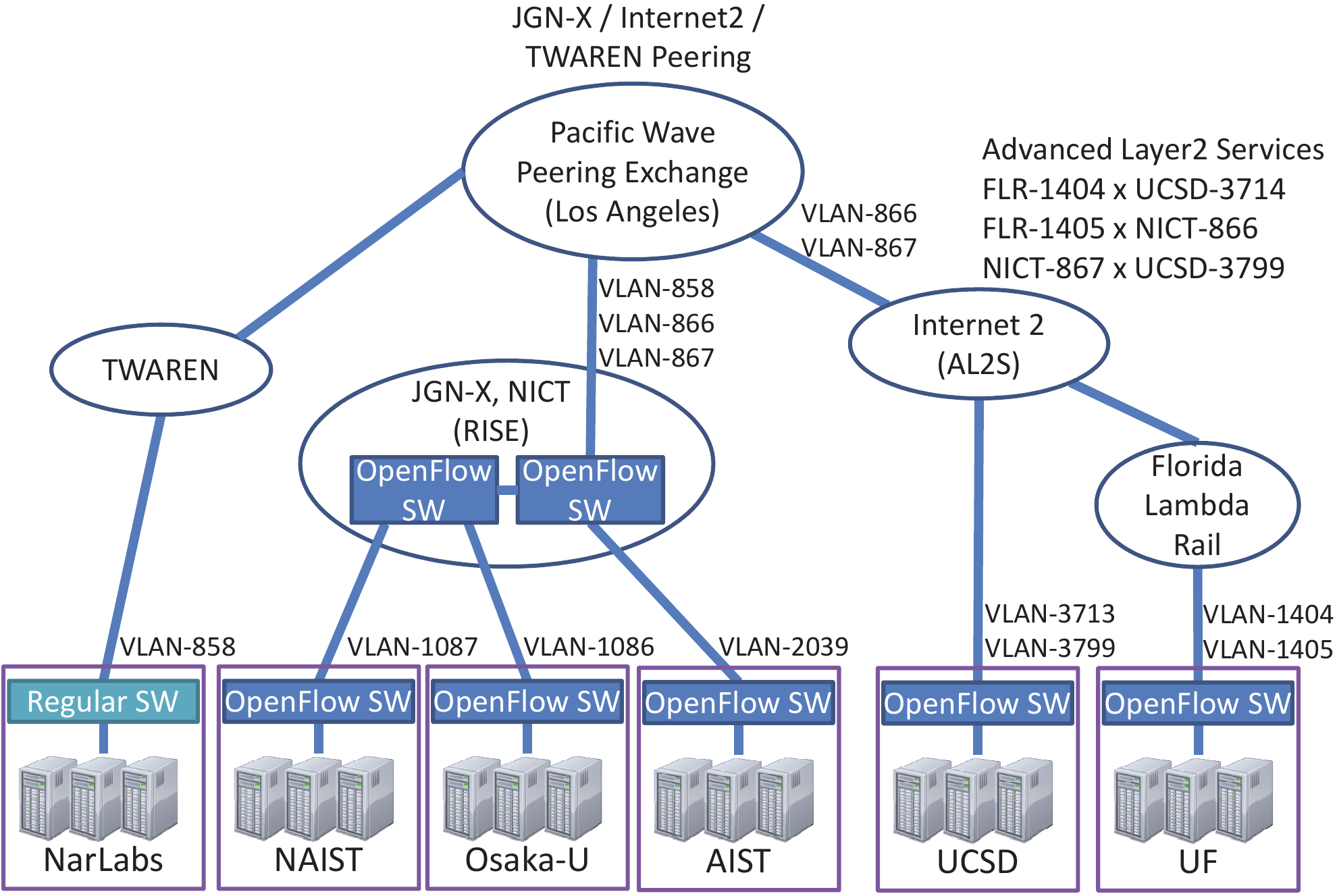}
	\caption{PRAGMA-ENT L2 Backbone. High-speed research networks (FLR, Internet2, JGN-X, TWAREN, and PacificWave) interconnects OpenFlow-enabled hardware switches in the USA (University of Florida and University of California San Diego), Japan (Nara Institute of Science and Technology, Osaka University, and National Institute of Advanced Industrial Science and Technology), and Taiwan (National Applied Research Laboratories).}
	\label{fig:ent-l2-backbone}
\end{figure}

PRAGMA-ENT team worked with Internet2, Florida Lambda Rail (FLR), National Institute of Information and Communications Technology (NICT), TWAREN, and Pacific Wave engineers to establish an international Layer-2 (reliable direct point-to-point data connection) backbone. Virtual Local Area Networks (VLANs) were allocated to create logical direct paths between (1) University of Florida (UF) and University of California San Diego (UCSD); (2) UF and NICT; (3) UCSD and NICT; and (4) National Applied Research Laboratories (NarLab) and NICT. Those paths from the US and Taiwan to Japan were expanded to reach Nara Institute of Science and Technology (NAIST), Osaka University and National Institute of Advanced Industrial Science and Technology (AIST). Associated technologies were used to create seamless connections between OpenFlow switches, deployed at participating sites (Figure \ref{fig:ent-l2-backbone}). In addition, GRE tunnel links over the Internet were also deployed as alternative paths. Since all OpenFlow switches are interconnected, independent of the geographical location, it is possible to develop SDN controllers to manage trust envelopes. Initial testing of the network achieved 1 Gbps throughput between the following direct paths, UF/UCSD, UF/NAIST, and UCSD/NAIST.

\subsubsection{Direct Wide-Area Layer-2 Path vs. GRE Tunneling}

\begin{table}[t]
  \caption{TCP thorughput via a direct Wide-Area L2 path, UCSD-UF (Mbps).}
  \label{table:thorughput-l2}
  \centering
  \begin{tabular}{c|ccccccc}
           & \multicolumn{7}{|c}{\# of threads} \\
  time (s) &   1 &   2 &   3 &   4 &   6 &   8 &   10 \\
  \hline
0	&   0	&   0	&   0	&   0	&   0	&   0	&   0 \\
5	& 213	& 393	& 540	& 594	& 598	& 632	& 693 \\
10	& 346	& 681	& 939	& 924	& 937	& 938	& 936 \\
15	& 343	& 682	& 938	& 911	& 938	& 921	& 944 \\
20	& 346	& 680	& 939	& 940	& 908	& 937	& 937 \\
25	& 346	& 680	& 939	& 937	& 939	& 934	& 934 \\
  \end{tabular}
\end{table}

\begin{table}[t]
  \caption{TCP throughput via GRE tunnel, UCSD-NAIST (Mbps).}
  \label{table:thorughput-gre}
  \centering
  \begin{tabular}{c|ccccccc}
           & \multicolumn{6}{|c}{\# of threads} \\
  time (s) &   1 &   2 &   3 &   4 &   8 &   10 \\
  \hline
0	&   0	&   0	&   0	&   0	&   0	&   0 \\
5	& 61.4	& 110	& 182	& 195	& 311	& 122 \\
10	& 38	& 121	& 183	& 244	& 317	& 178 \\
15	& 27.6	& 95	& 197	& 222	& 307	& 379 \\
20	& 30	& 99.6	& 197	& 230	& 343	& 669 \\
25	& 34.5	& 106	& 260	& 267	& 533	& 675 \\
30	& 41.5	& 133	& 265	& 278	& 628	& 650 \\
  \end{tabular}
\end{table}

In order to compare the performance of direct paths and tunnels, TCP throughput experiments were executed using resources connected through Internet2/FLR (UCSD and UF), and resources connected through GRE tunnels (NAIST and UCSD). Since machines are connected to OpenFlow switches via 1 Gbps links, the maximum throughput is 1 Gbps. The table \ref{table:thorughput-l2} and \ref{table:thorughput-gre} summarize the results. The operating system's network stack parameters were kept with default values. Multiple simultaneous TCP streams were used to push network links to its capacity. Note that the software processing overheads in a GRE tunnel makes it difficult to achieve high throughput. Even so, the performance of the GRE links has reached around 650 Mbps; they are still useful as alternative paths.

\subsection{Control Plane}
PRAGMA-ENT is architected to provide a virtual network slice for each application, user, and/or project in order to enable independent and isolated development and evaluation of software-defined network functions. This architecture allows the participating researchers to share a single OpenFlow network backbone with multiple network experiments simultaneously, and gives them freedom to control the network resources. For this purpose, PRAGMA-ENT has evaluated different technologies for the control plane, such as FlowVisor~\cite{sherwood2009flowvisor}, OpenVirteX~\cite{Al-Shabibi2014a}, and FlowSpace Firewall~\cite{fsfw}. As a result, the slicing technology AutoVFlow~\cite{Yamanaka2014} will be used to create virtual network slices.

\subsubsection{AutoVFlow}
In order to support multiple experiments (i.e. enable multiple OpenFlow controllers to co-exist), AutoVFlow is a suitable approach on the PRAGMA-ENT infrastructure. There are OpenFlow network virtualization techniques other than AutoVFlow. FlowSpace Firewall is a VLAN-based OpenFlow network virtualization technique, however, it requires the configuration of VLANs in the infrastructure and the distributed management of PRAGMA-ENT makes this approach not practical. Also, we have already used VLANs to construct the L2 backbone of the network; and using VLANs over VLANs is also not practical because not all switches can speak nested VLAN. OpenVirteX does not require multiple VLANs in the L2 backbone network. Instead, OpenVirteX implements virtualization by IP address space division and translation for virtual OpenFlow networks. In order to isolate flows of different virtual OpenFlow networks, OpenVirteX divides the IP address space for each virtual OpenFlow network. To support conflicting IP addresses in different virtual OpenFlow networks, OpenVirteX translates them to the IP addresses in the divided space. In the OpenVirteX architecture, a single proxy is responsible for the management of the IP address division and translation that affect all OpenFlow switches in the network, making this approach not well suitable for PRAGMA-ENT.

\begin{figure}
	\centering
	\includegraphics[width=\linewidth]{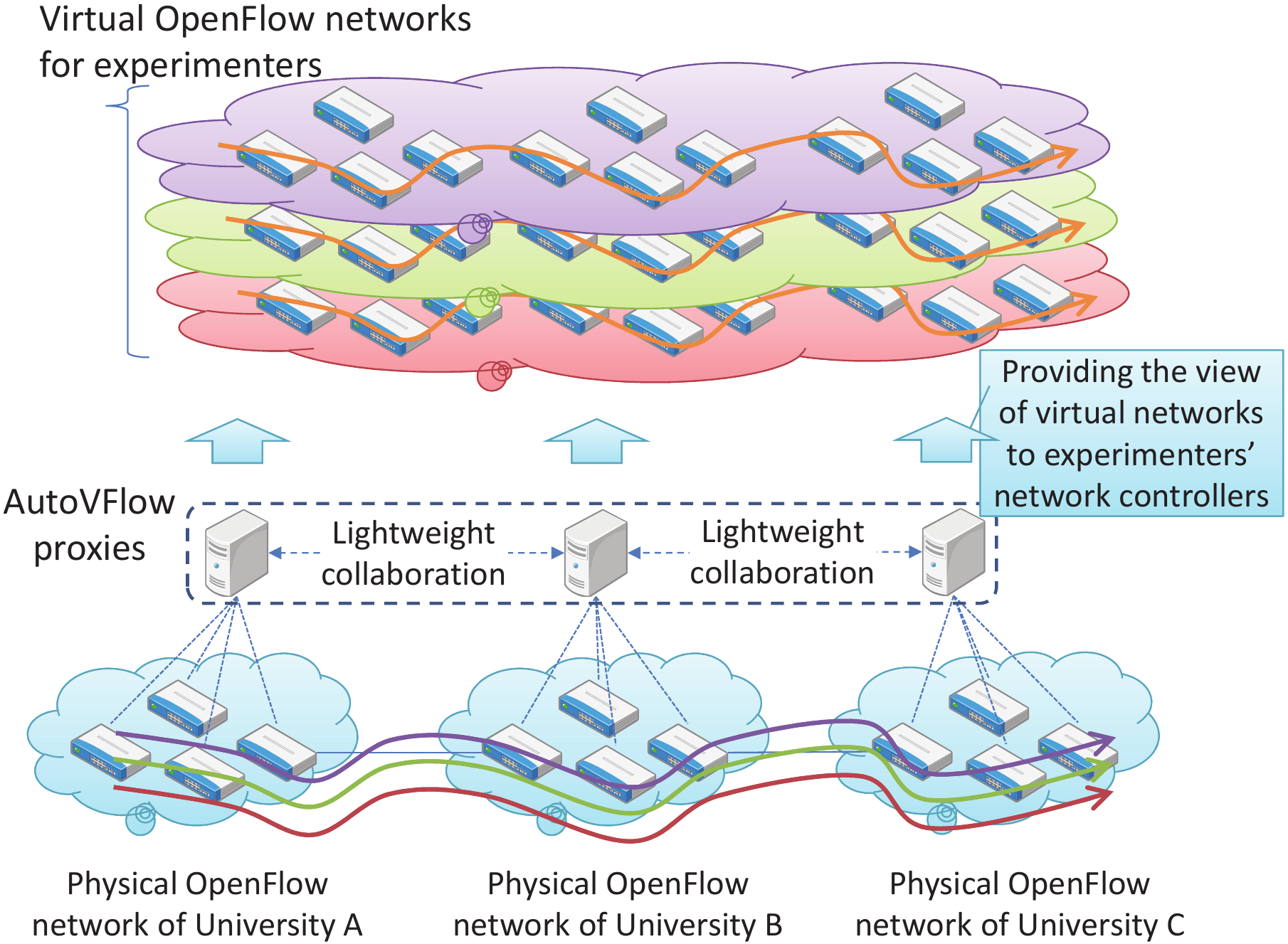}
	\caption{Implementation and deployment of AutoVFlow. Multiple AutoVFlow proxies are deployed for each administrative domain instead of using a central slicing controller to avoid a single point of failure.}
	\label{fig:autovflow}
\end{figure}

AutoVFlow addresses the shortcomings discussed above of FlowSpace Firewall and OpenVirteX. Two unique features of AutoVFlow are that 1) it provides autonomous implementation of virtualization and topologies for virtualization, and also 2) provides automatic federation of different domains of OpenFlow networks. As illustrated in Figure 3, AutoVFlow is a virtual network slicing technology similar to FlowSpace firewall and OpenVirtex, however it provides a distributed implementation of slicing by controllers of different domains, instead of using a central slicing controller. In AutoVFlow architecture, multiple proxies are deployed for each administrative domain; and each proxy autonomously manages the address space division and translation for the part of the OpenFlow switches where the proxy has a responsibility to manage. When data packets are transferred between OpenFlow switches handled by different proxies, the proxy of the source switch automatically modifies data packet headers to be understood by the destination proxy.  Using AutoVFlow, each site administrator can operate AutoVFlow proxies autonomously, and there is no single point of failure of the PRAGMA-ENT infrastructure. Hence, this architecture is more appropriate for a widely distributed network testbed composed of different administrative domains as is the case in PRAGMA-ENT.

\section{Applications on PRAGMA-ENT}

PRAGMA-ENT is a very unique global scale OpenFlow network. It accepts user OpenFlow controllers; and the participating researchers can deploy their own OpenFlow controller and perform their network experiments leveraging the global scale OpenFlow network. Several applications and experiments have been evaluated on PRAGMA-ENT. This section introduces some of the applications.

\subsection{Bandwidth and latency aware routing}

\begin{figure}
	\centering
	\includegraphics[width=\linewidth]{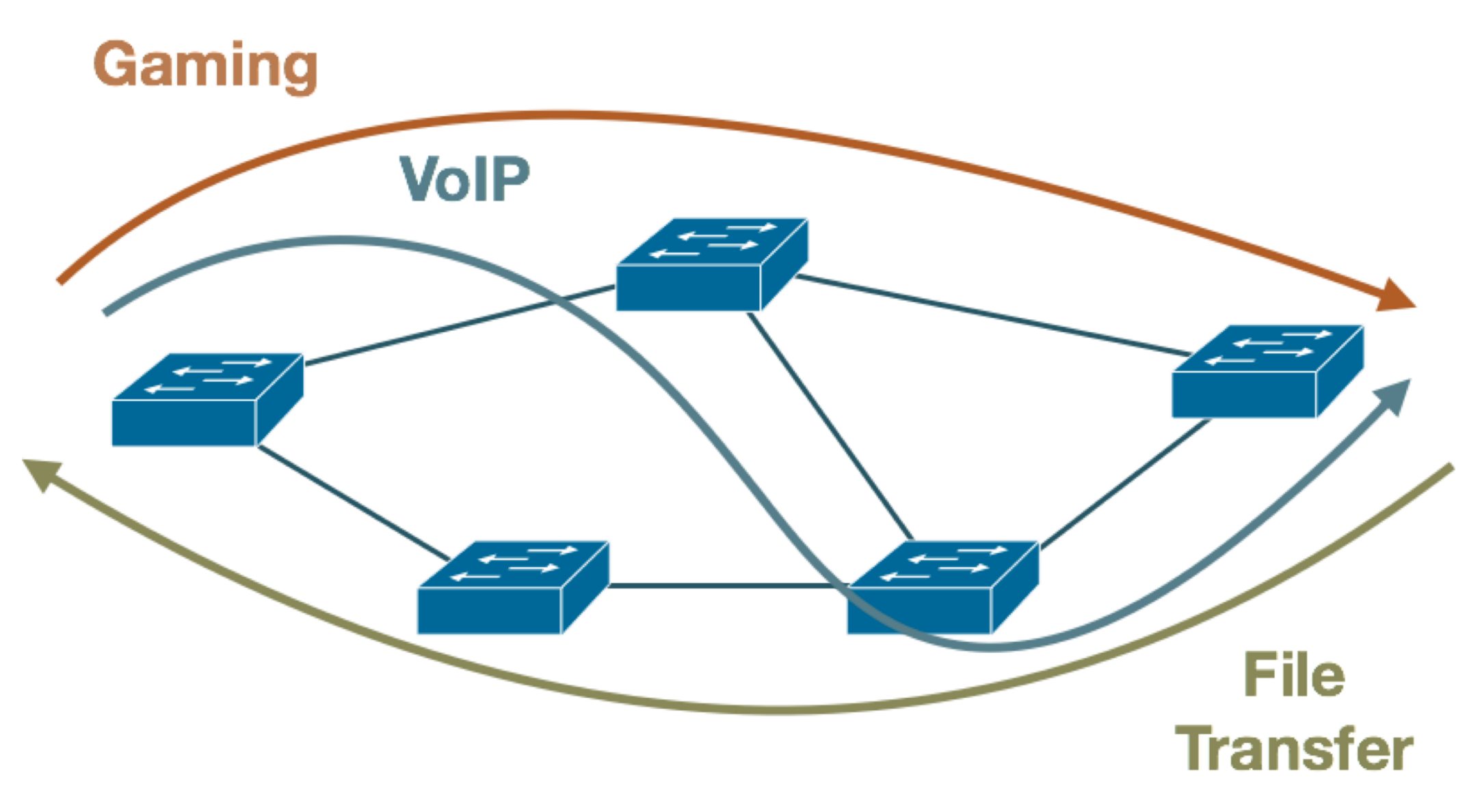}
	\caption{Concept of bandwidth and latency aware routing.}
	\label{fig:overseer}
\end{figure}

Bandwidth and latency aware routing is a routing concept proposed in the papers by Pongsakorn et al. \cite{U-chupala2014} and Ichikawa et al. \cite{Ichikawa2013}. Its core idea is to dynamically optimize the route of each flow according to whether the flow is bandwidth-intensive or latency-oriented. Figure \ref{fig:overseer} illustrates this concept. In the distributed wide area network environment, the bandwidth and latency vary according to the paths; and the optimal path is different for each of applications.

\begin{figure}
	\centering
	\includegraphics[width=\linewidth]{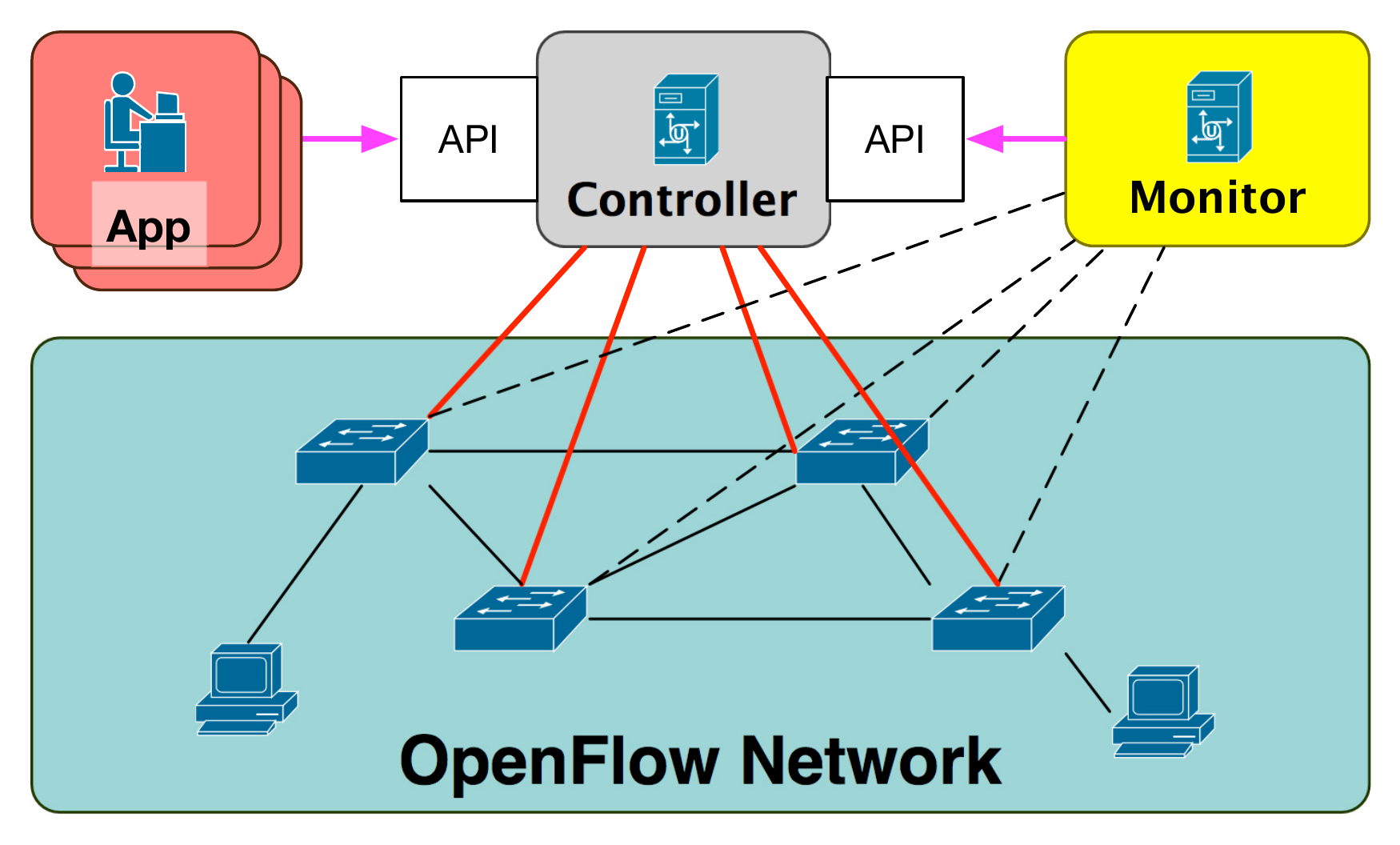}
	\caption{Architecture and structure of Overseer.}
	\label{fig:overseer_architecture}
\end{figure}

Overseer is an implementation of bandwidth and latency aware routing as OpenFlow controller. It comprises of 4 primary components; OpenFlow network, OpenFlow controller, network monitor and supported application. Communication between components is done through a set of APIs to allow each component to be easily replaceable. Figure \ref{fig:overseer_architecture} shows the relationship and information flow direction between each component.

Overseer was deployed over the entire PRAGMA-ENT to evaluate its practicality. The evaluation was done by measuring actual bandwidth and latency of flows in the network with Overseer's dynamic routing compare with traditional shortest-path routing. The results of the evaluation showed that Overseer improved network performance significantly in terms of both bandwidth and latency.

\subsection{Multipath routing}

In the PRAGMA-ENT, there are several possible paths from one site to another. Multipath routing is a routing technology improving the bandwidth by aggregating available bandwidth from those multiple paths between source and destination. We have been evaluating two types of multipath routing: application level and network transport level. 

\subsubsection{Multipath GridFTP}

\begin{figure}[t]
\centering
\includegraphics[width=\linewidth]{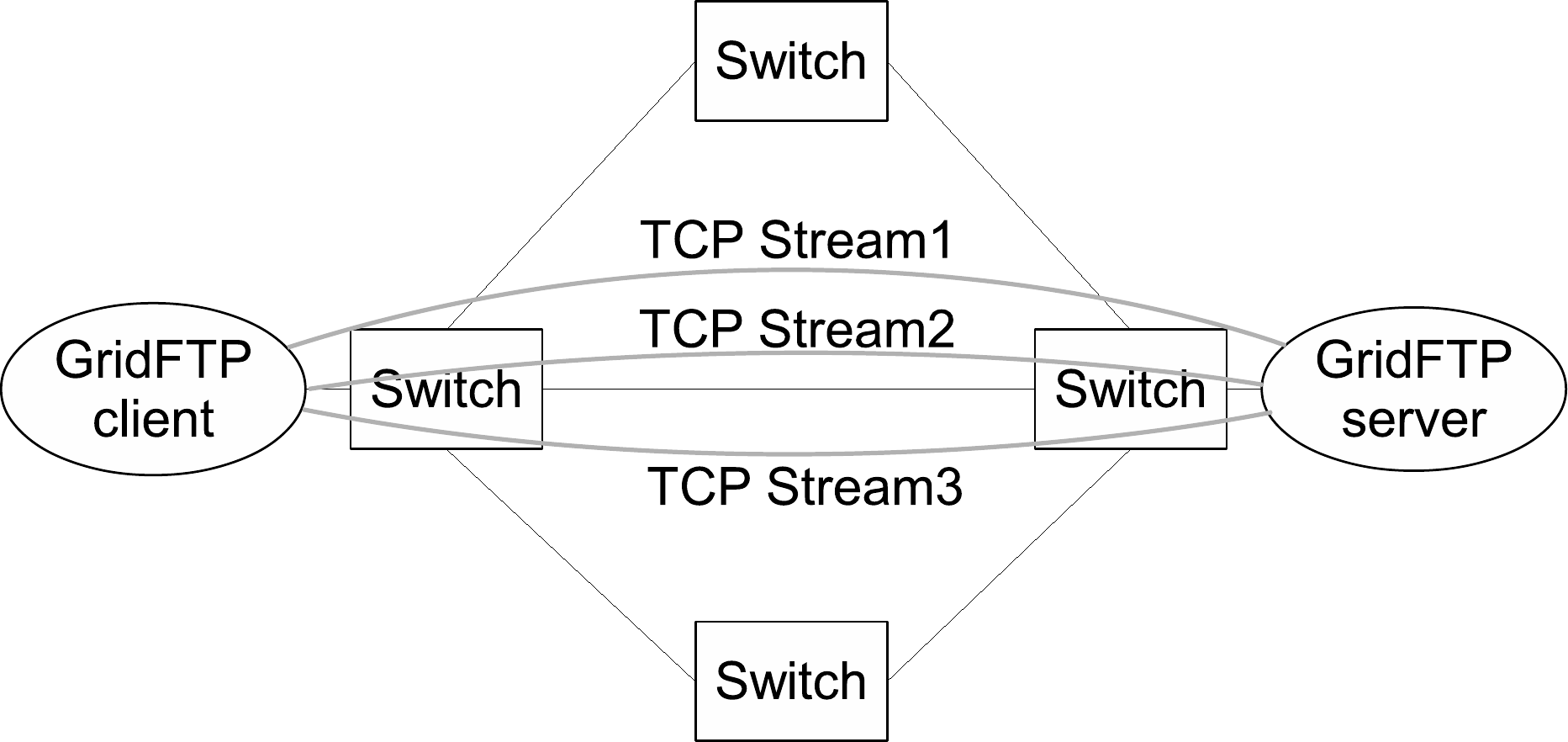}
\caption{Parallel transfer of the conventional GridFTP.}
\label{fig:usual-GridFTP}
\end{figure}

\begin{figure}[t]
\centering
\includegraphics[width=\linewidth]{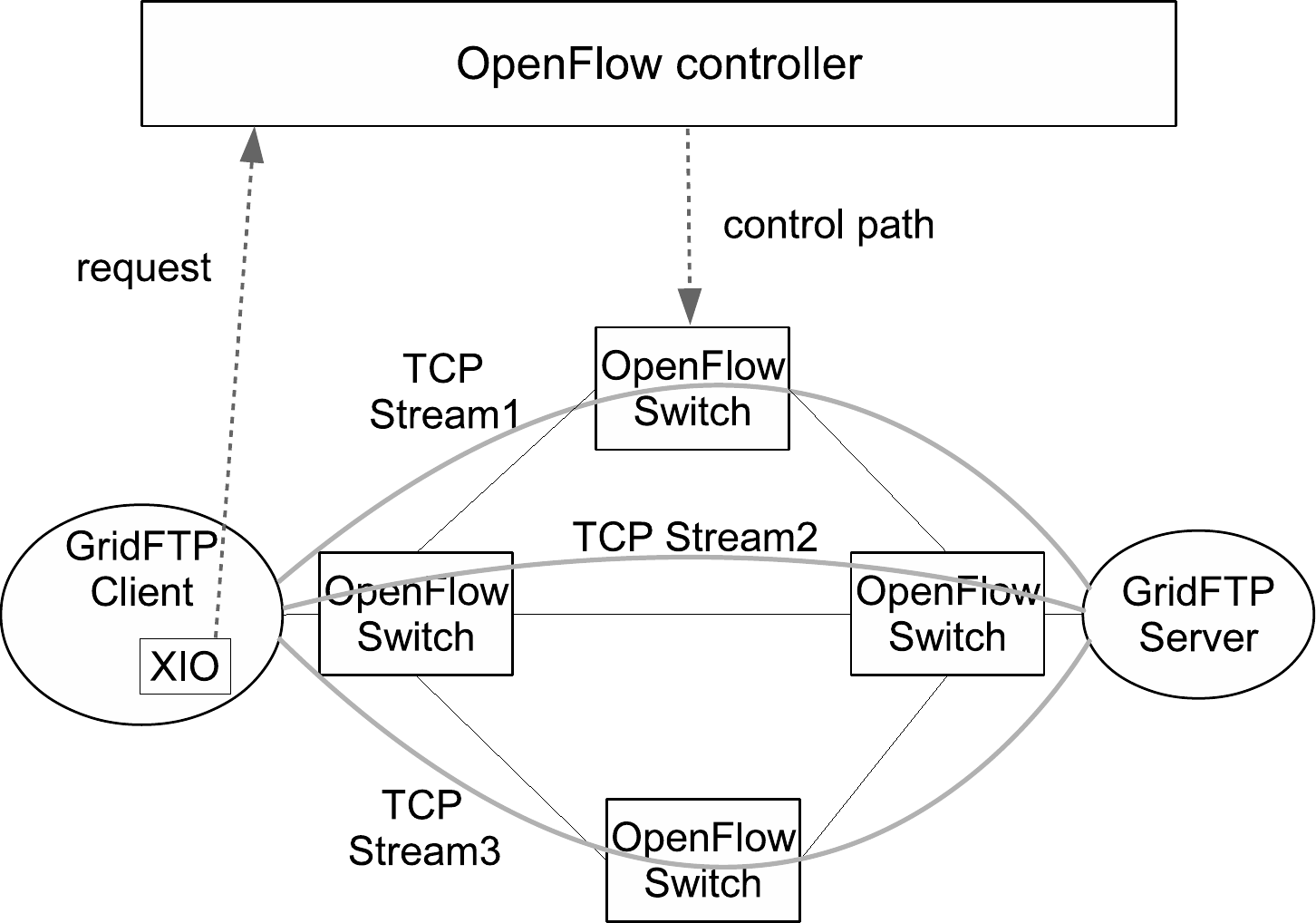}
\caption{Parallel transfer of Multipath GridFTP.}
\label{fig:propose-SDN-GridFTP}
\end{figure}

\begin{figure}[t]
\centering
\includegraphics[width=\linewidth]{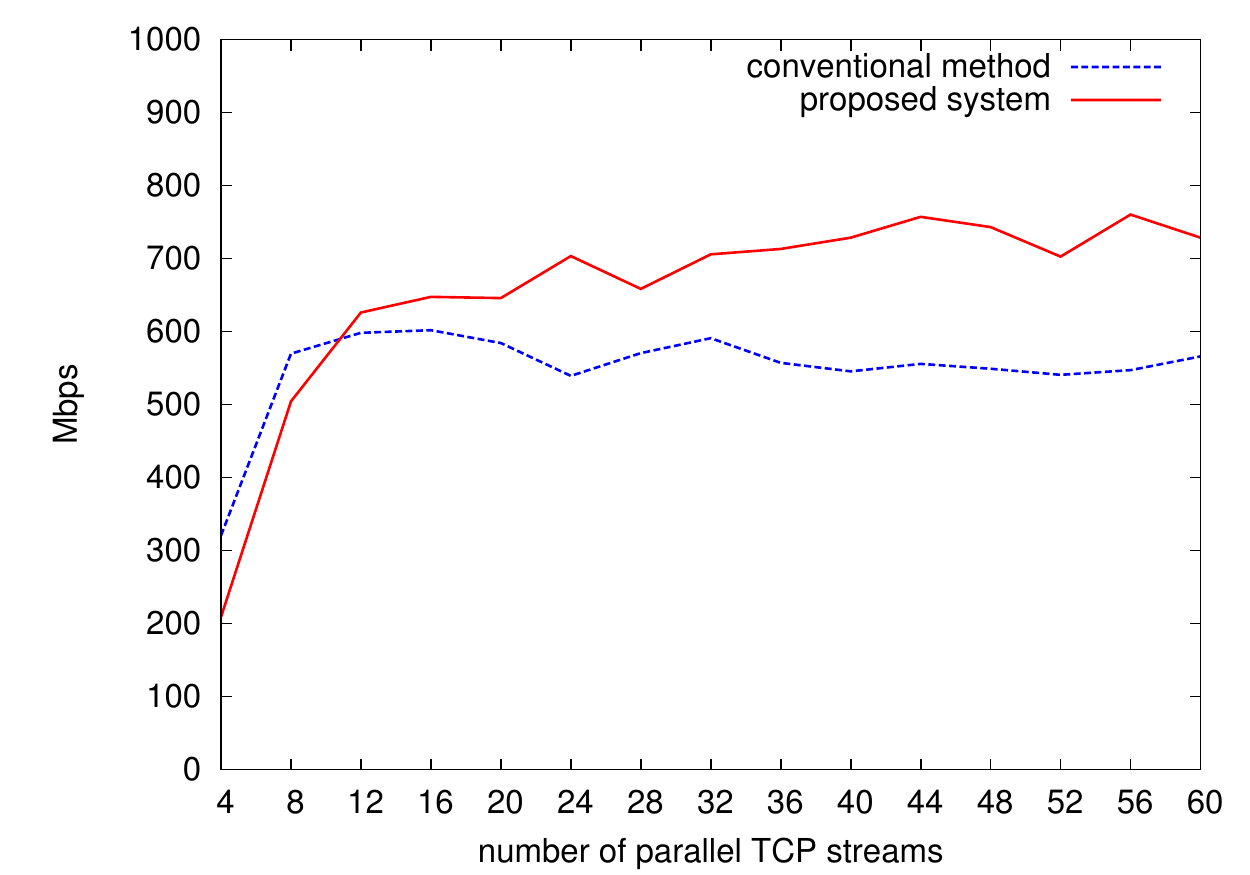}
\caption{Average data transfer bandwidth of the proposed system and the conventional method.}
\label{fig:GridFTP-result}
\end{figure}

GridFTP supports a parallel data transfer scheme by using multiple TCP streams on application level to realize high speed transfer between sites \cite{foster2001anatomy,GlobusOnline}, and it was widely used in the field of Grid computing. Figure \ref{fig:usual-GridFTP} shows the parallel transfer of the conventional GridFTP. The conventional GridFTP basically takes only a single shortest path even if there are available multiple paths, because multiple TCP streams by GridFTP are routed according to the default IP routing protocol. On the other hand, as shown in Figure \ref{fig:propose-SDN-GridFTP}, our multipath GridFTP distributes the parallel TCP streams of GridFTP into multiple network paths. To control the distribution of the multiple TCP streams, we have designed a OpenFlow controller for multipath GridFTP \cite{Huang2015}. In our system, a client requests to the controller to assign multiple available paths in advance of the actual data transfer, then the TCP streams from the client are distributed to different paths by the controller.

Multipath GridFTP system was also deployed over the entire PRAGMA-ENT to evaluate its practicality. The evaluation of Multipath GridFTP system was performed by measuring the actual transfer time and used bandwidth of each TCP stream in the network. Figure \ref{fig:GridFTP-result} shows the average data transfer bandwidth using four different paths from NAIST to UF. As shown in the results, our proposed method has achieved better performance using four paths while the conventional method uses just a single shortest path.

\subsubsection{Multipath TCP}

\begin{figure}
  \centering
  \includegraphics[width=\linewidth]{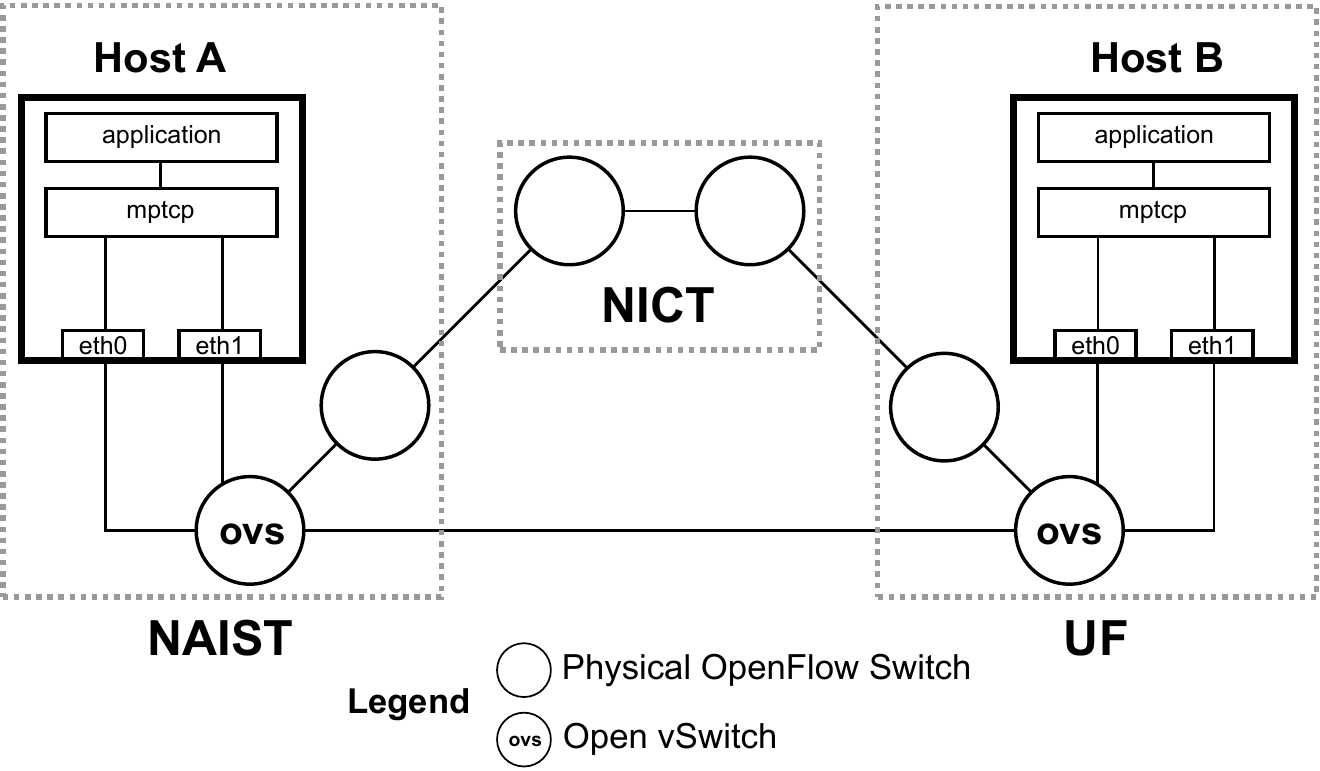}
  \caption{Illustration of the Multipath TCP routing experiment on PRAGMA-ENT.}
  \label{fig:mptcp_ent}
\end{figure}

\begin{figure}
  \centering
  \includegraphics[width=\linewidth]{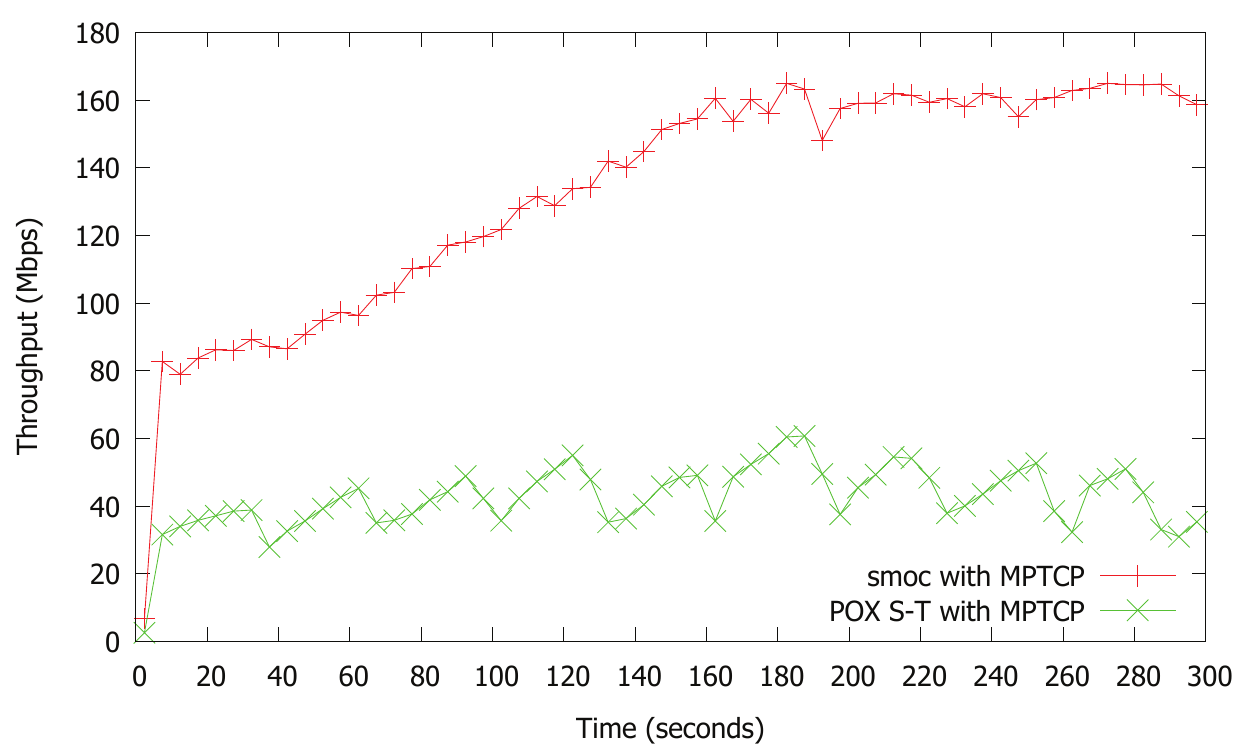}
  \caption{Bandwidth between two hosts measured by iperf on PRAGMA-ENT.}
  \label{fig:plot_ent}
\end{figure}

Multipath TCP (MPTCP) \cite{rfc6182,rfc6824} is a widely-researched mechanism that allows an application to create more than one TCP stream in one network socket. While having more than one stream can be beneficial to TCP, guaranteeing that those streams use different paths with minimal conflict may lead to better performance. Also, unlike application level multiple TCP streams, handling those multiple TCP streams are implemented behind the socket library and it does not require any modification to the application.

The Simple Multipath OpenFlow Controller (smoc) project, based on heavily modified Overseer code supported by POX, was started to create a simple, primarily topology-based controller that performs this task. When an MPTCP instance is established using MP\_CAPABLE option, smoc creates a path set, which is a collection of multiple paths between a pair of hosts, and associates the path set to the instance. The path set is ranked topologically, mostly preferring the shortest paths. When subsequent subflows are established using MP\_JOIN option, smoc uses the next path in the path set. In this way, MPTCP traffic can be distributed to multiple paths and more bandwidth is available to the MPTCP instance.

smoc was tested on a section of PRAGMA-ENT to evaluate its performance in the wide-area network. The deployment of the testbed is illustrated in Figure \ref{fig:mptcp_ent}. It was compared to POX's spanning-tree default example controller. As shown in Figure \ref{fig:plot_ent}, we achieved satisfactory results as smoc could provide multipath routing to MPTCP-enabled hosts without adversely affecting hosts without MPTCP.

\subsection{eScience visualization}

Visualization in eScience applications relies on the network of a distributed environment. Thus, where scientists view the computational results is geographically different compared to where the data was processed; and the processed results need to be moved over the network. Improvment of network usability, performance, reliability and efficiency will solve some of the network issues that cause problems for remote visualization.

\subsubsection{Satellite Image Sharing between Taiwan and Japan}

\begin{figure}
  \centering
  \includegraphics[width=\linewidth]{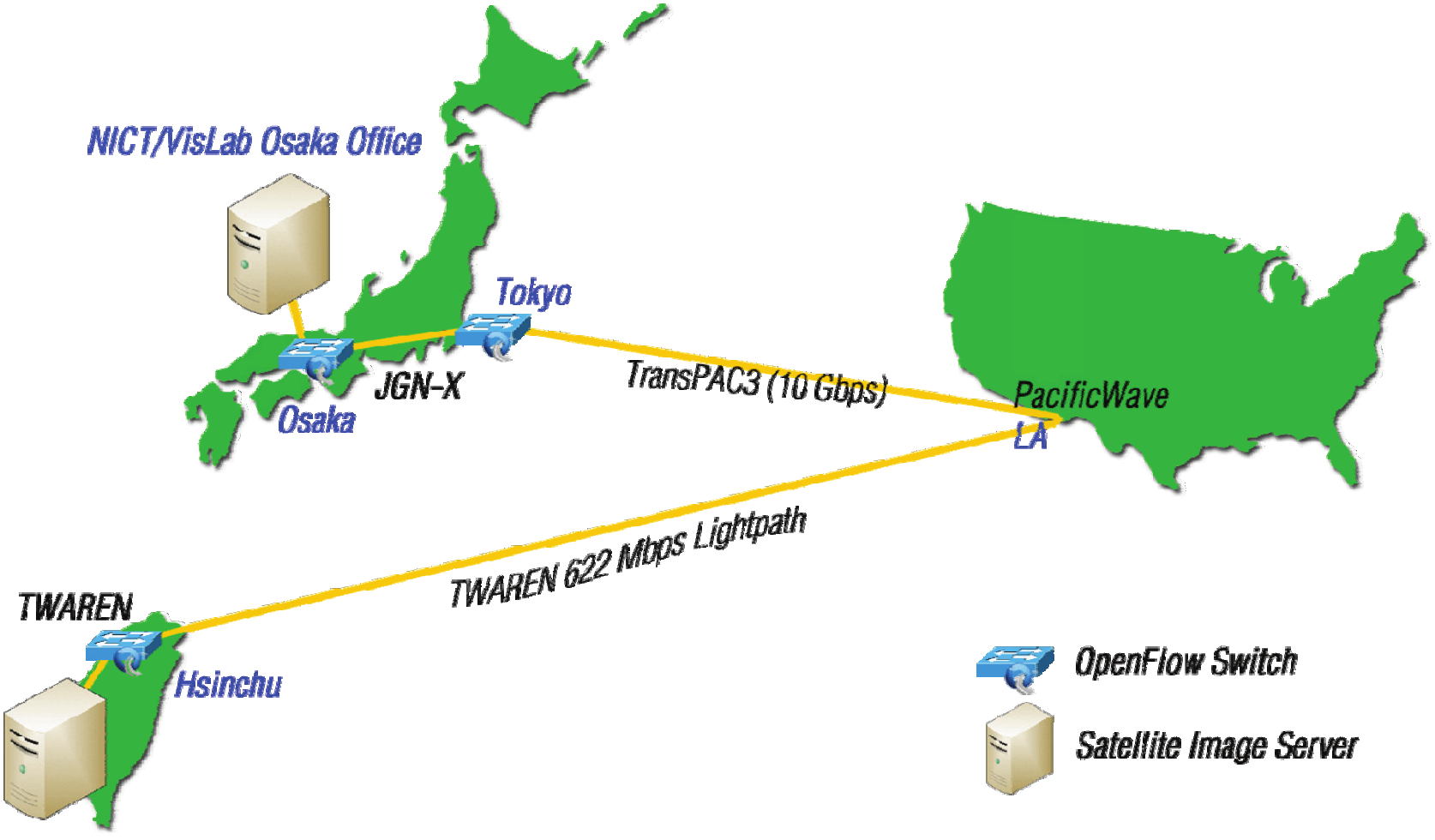}
  \caption{Satellite image sharing emvironment on PRAGMA-ENT.}
  \label{fig:satellite-tw-jp}
\end{figure}

Satellite image sharing between Taiwan and Japan is a newly emerging international project in rapid response to natural disasters, such as tsunami, earthquake, flood and landslide. Emergency observation, near real-time satellite image processing and visualization are critical for supporting the responses to these types of natural disasters. In case of an emergency, we can make a request for emergency observation to satellite image services. Those satellite image services will respond to the request by sending the satellite images within a couple hours, and these observed satellite images need to be transferred to computational resources for image processing and visualization. Therefore, an on-demand high-speed network is very important for rapid disaster responses.

This project created an on-demand network on the PRAGMA-ENT with a virtual network slice on the AutoVFlow architecture. Currently, Taiwan and Japan are connected via the United States on PRAGMA-ENT (Figure \ref{fig:satellite-tw-jp}). Our preliminary evaluations on transferring data from Taiwan to Japan via US indicate that we acheive more than double the speed compared to just using the Internet between Taiwan and Japan. In the future, we plan to use more efficient data transfer mechanisms such as multipath routing mentioned in the previous sections to improve the performance further.

\subsubsection{Flow Control for Streamings on Tiled Display Wall}

\begin{figure}
  \centering
  \includegraphics[width=\linewidth]{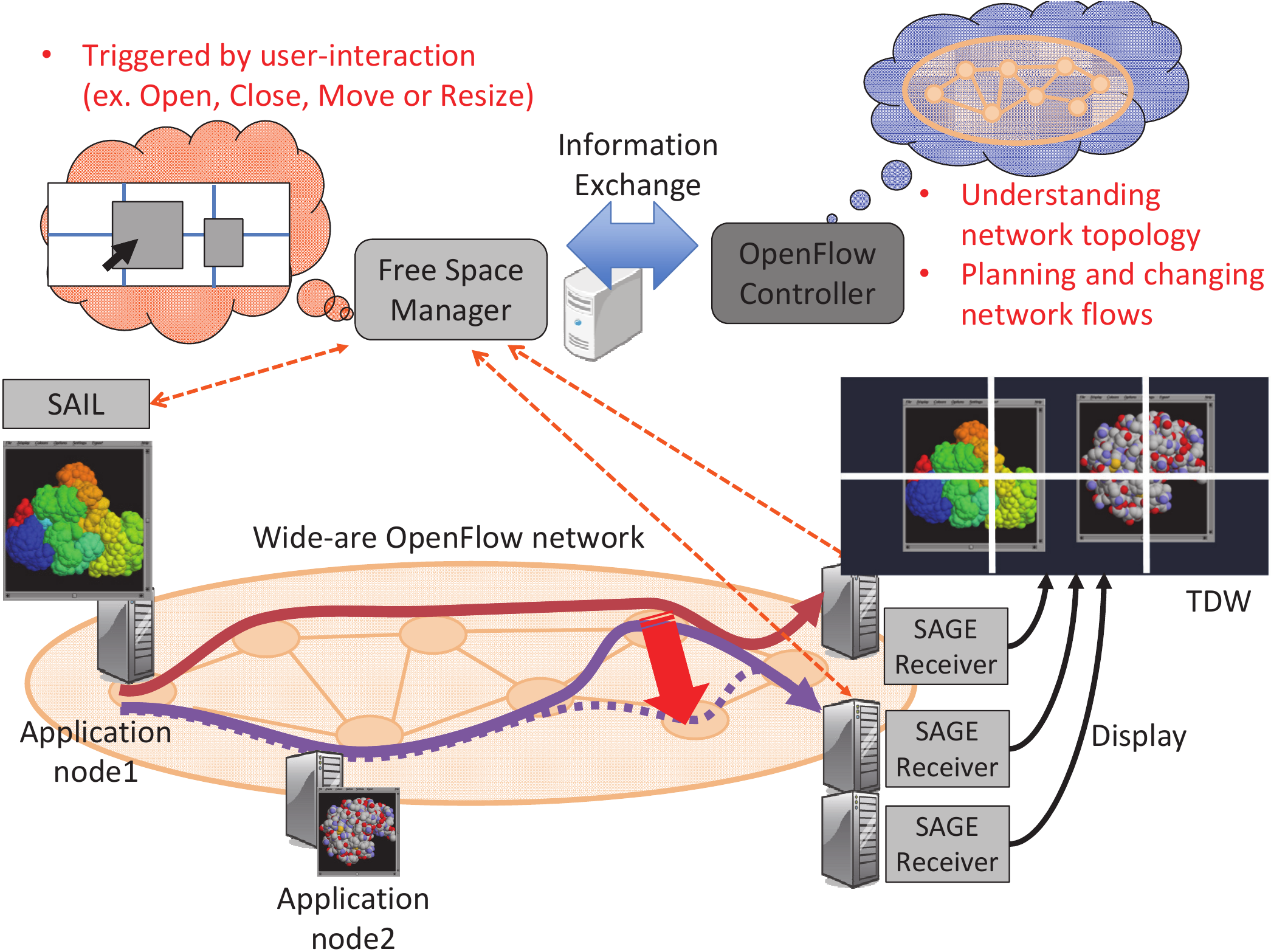}
  \caption{Overview of Flow Control for Streamings on Tiled display wall.}
  \label{fig:sage-sdn}
\end{figure}

A large amount of scientific data needs to be visualized and then shared among scientists over the Internet for scientific discovery and collaboration. Tiled Display Wall (TDW) is a set of display panels arranged in a matrix for high-resolution visualization of large-scale scientific data. Scalable Adaptive Graphics Environment (SAGE) is a TDW middleware, which is increasingly used by scientists who must collaborate on a global scale for problem solving~\cite{SAGE}. The reason is that SAGE-based TDW allows scientists to display multiple sets of visualized data, each of which is located on a geographically-different administrative domain, in a single virtual monitor. However, SAGE does not have any effective dynamic routing mechanism for packet flows, despite that SAGE heavily relies on network streaming technique to visualize remote data on a TDW. Therefore, user-interaction during visualization on SAGE-based TDW sometimes results in network congestion and as a result leads to a decrease of visualization quality caused by a decrease in frame rate. 

For the reason above, we proposed and developed a dynamic routing mechanism that switches packet flows onto network links where better performance is expected, in response to user-interaction such as window movement and resizing (Figure \ref{fig:sage-sdn}). Technically, we have leveraged OpenFlow, an implementation of Software Defined Networking, to integrate network programmability into SAGE. At Supercomputer 2014 (SC14), our research team showed that SAGE enhanced with the proposed mechanism mitigates the network congestion and improves the quality of visualization on the TDW over the wide area OpenFlow network on the Internet~\cite{INDIS2014}.

\section{Summary and Future Plans}
In closing, we have established a network testbed for use by different Pacific Rim researchers and institutes that are part of the PRAGMA community. The network testbed offers complete freedom for researchers to access network resources without concerns of interfering with a production network. Our future plans include the expansion of the network to include more sites in the Pacific Rim area and establish a more persistent testbed that is available for use. In particular, ViNe overlays will be deployed to expand PRAGMA-ENT to sites without direct connection to the backbone (e.g., commercial clouds such as Azure \cite{azure}). This semi-permanent section of PRAGMA-ENT will utilize AutoVFlow and FlowSpace Firewall as core technologies to enable the creation of experimental network slices. We will also plan to deploy perfSONAR in order to monitor the testbed infrastructure during experiments. Another key component to be added will be better tools for end user support, including easy-to-use scheduling UI and easier user management for administrators. The management and monitoring components will be housed in a PRAGMA-ENT operations center.
 
Once the infrastructure is established, it will be tested with several use-cases, including executing molecular simulations using DOCK and use the SDN monitoring capabilities to profile communication pattern during DOCK execution, and using SDN to address data licensing and security for the biodiversity mapping tool LifeMapper (http://lifemapper.org).

%ACKNOWLEDGMENTS are optional
\section{Acknowledgments}
This work was supported in part by PRAGMA (NSF OCI 1234983), UCSD PRIME Program, the collaborative research of NICT and Osaka University (Research on high functional network platform technology for large-scale distributed computing), JSPS KAKENHI (15K00170) and a research award from Microsoft Azure4Research. Any opinions, findings and conclusions or recommendations expressed in this material are those of the authors and do not necessarily reflect the views of NSF and/or Microsoft.

\bibliographystyle{abbrv}
\bibliography{references}  % sigproc.bib is the name of the Bibliography in this case
% ACM needs 'a single self-contained file'!
%

\end{document}